\documentclass[
notoc,nohyper]{JHEP} 

\usepackage[psamsfonts]{amsfonts}

\newcommand{\tr}{\mathop{\rm tr}\nolimits}
\newcommand{\Tr}{\mathop{\rm Tr}\nolimits}
\newcommand{\SU}{\mathop{\rm SU}}
\newcommand{\U}{\mathop{\rm {}U}}
\newcommand{\rmd}{{\rm d}}

\newcommand\fverb{\setbox\pippobox=\hbox\bgroup\verb}
\newcommand\fverbdo{\egroup\medskip\noindent%
			\fbox{\unhbox\pippobox}\ }
\newcommand\fverbit{\egroup\item[\fbox{\unhbox\pippobox}]}
\newbox\pippobox
\title{CP breaking in lattice chiral gauge theories}

\author{Kazuo Fujikawa\\
Department of Physics, University of Tokyo, Bunkyo-ku, Tokyo 113, Japan}
\author{Masato Ishibashi\\
Department of Physics, University of Tokyo, Bunkyo-ku, Tokyo 113, Japan}
\author{Hiroshi Suzuki\\
Department of Mathematical Sciences, Ibaraki University, Mito 310-8512,
Japan\\
E-mail: \email{hsuzuki@mito.ipc.ibaraki.ac.jp}}
\received{\today} 		
\accepted{\today}		

\preprint{UT-02-15\\IU-MSTP/47
\\\heplat{0203016}}	

\abstract{
The CP symmetry is not manifestly implemented for the local and
doubler-free Ginsparg-Wilson operator in lattice chiral gauge theory. We
precisely identify where the effects of this CP breaking appear. We show
that they appear in: (I)~Overall constant phase of the fermion
generating functional. (II)~Overall constant coefficient of the fermion
generating functional. (III)~Fermion propagator appearing in external
fermion lines and the propagator connected to Yukawa vertices. The first
effect appears from the transformation of the path integral measure and
it is absorbed into a suitable definition of the constant phase factor
for each topological sector; in this sense there appears no
``CP anomaly''. The second constant arises from the explicit breaking in
the action and it is absorbed by the suitable weights with which
topological sectors are summed. The last one in the propagator is
inherent to this formulation and cannot be avoided by a mere
modification of the projection operator, for example, in the framework
of the Ginsparg-Wilson operator. This breaking emerges as an (almost)
contact term in the propagator when the Higgs field, which is treated
perturbatively, has no vacuum expectation value. In the presence of the 
vacuum expectation value, however, a completely new situation arises and
the breaking becomes intrinsically non-local, though this breaking may
still be removed in a suitable continuum limit. This non-local CP
breaking is expected to persist for a non-perturbative treatment of the
Higgs coupling.
}


\begin{document}

\maketitle 

\section{Introduction}
Discovery of gauge covariant local lattice Dirac
operators~\cite{Hasenfratz:1998ft,Neuberger:1998fp}, which satisfy the
Ginsparg-Wilson relation~\cite{Ginsparg:1982bj}, paved a way to a
manifestly local and gauge invariant lattice formulation of anomaly-free
chiral gauge theories~\cite{Luscher:1999du}--\cite{Kikukawa:2001mw}.%
\footnote{The locality of the overlap Dirac
operator~\cite{Neuberger:1998fp} was shown in
refs.~\cite{Hernandez:1999et,Neuberger:2000pz}.} See also related early
works in refs.~\cite{Narayanan:1993wx,Randjbar-Daemi:1995sq}, and
refs.~\cite{Niedermayer:1999bi}--\cite{Kikukawa:2001jk} for reviews.

It has been however pointed out that the CP symmetry, the fundamental
discrete symmetry in chiral gauge theories, is not manifestly
implemented in this formulation. The basic observation related to this
effect is as follows~\cite{Hasenfratz:2001bz}: In the formulation, the
chirality is imposed through~\cite{Narayanan:1998uu,Niedermayer:1999bi}
\begin{equation}
   P_-\psi=\psi,\qquad\overline\psi\overline P_+=\overline\psi,
\label{eq:onexone}
\end{equation}
where $\overline P_\pm=(1\pm\gamma_5)/2$ and
$P_\pm=(1\pm\hat\gamma_5)/2$ with $\hat\gamma_5=\gamma_5(1-2aD)$, and
$D$~is the Dirac operator. The Ginsparg-Wilson relation then guarantees
that the kinetic term is consistently decomposed,
$\overline\psi\overline P_+D\psi=\overline\psi D P_-\psi$. However,
since the above condition is not symmetric in the fermion and
anti-fermion and since CP exchanges these two, CP symmetry is explicitly
broken. In fact, the fermion action in the case of pure chiral gauge
theory without Higgs couplings changes under CP as\footnote{For our
conventions, see Appendix~A. Note that the action of vector-like gauge
theories is manifestly invariant under this CP transformation.}
\begin{equation}
   S_{\rm F}=a^4\sum_x\overline\psi(x)\overline P_+D P_-\psi(x)
   \to S_{\rm F}=a^4\sum_x\overline\psi(x)\gamma_5P_+
   \gamma_5D\overline P_-\psi(x),
\end{equation}
and this causes the change in the propagator 
\begin{eqnarray}
   &&{\langle\psi(x)\overline\psi(y)\rangle_{\rm F}
   \over\langle1\rangle_{\rm F}}
   =P_-{1\over D}\overline P_+(x,y),
\nonumber\\
   &&\to{\langle\psi(x)\overline\psi(y)\rangle_{\rm F}
   \over\langle1\rangle_{\rm F}}
   =\overline P_-{1\over D}\gamma_5 P_+\gamma_5(x,y)
   =P_-{1\over D}\overline P_+(x,y)
   -a\gamma_5a^{-4}\delta_{x,y}.
\end{eqnarray}

One might think that a suitable modification of the chiral projectors
would remedy this CP breaking. That this is impossible with the standard
CP transformation law has been shown~\cite{Fujikawa:2002is} under rather
mild assumptions for a general class of Ginsparg-Wilson operators, as
long as they are local and free of species doublers. The above CP
breaking is thus regarded as an inherent feature of this formulation.
Also, the chirality constraint~(\ref{eq:onexone}) is very fundamental
and it influences the construction of the fermion integration measure,
one might then worry that the CP breaking emerges in many other places
and an analysis of CP violation (in the conventional sense) with this
formulation would be greatly hampered. It is thus important to precisely
identify where the effects of the above CP breaking inherent in this
formulation appear.

The purpose of this paper is to clarify the above issue. Following the
formulation of refs.~\cite{Luscher:1999du,Luscher:2000un}, we show that
the effects of CP violation  emerge only at: (I)~Overall constant phase
of the fermion generating functional. (II)~Overall constant coefficient
of the fermion generating functional. (III)~Fermion propagator appearing
in external fermion lines and the propagator connected to 
Yukawa vertices. Our result is summarized in
eq.~(\ref{eq:fourxtwentyone}) for pure chiral gauge theory and, when
there is a Yukawa coupling, in eq.~(\ref{eq:sixxeight}). The first two
constants above depend only on the topological sectors concerned and the
problem is reduced to the choice of weights with which various
topological sectors are summed. This problem, which is not particular to
this formulation, is thus fixed by a suitable choice of weight factors.
The last effect in the fermion propagator, on the other hand, is
inherent to this formulation. When the Higgs field has no expectation
value, as we have seen above, it emerges as an (almost) contact term.
However, in the presence of the Higgs expectation value, this breaking
becomes intrinsically non-local.

\section{Formulation}

\subsection{General Ginsparg-Wilson relation}
In our analysis, we assume that the Dirac operator satisfies a general
form of the Ginsparg-Wilson relation. In terms of the hermitian
operator~$H$
\begin{equation}
   H=a\gamma_5D=aD^\dagger\gamma_5=H^\dagger,
\label{eq:twoxone}
\end{equation}
the relation we adopt is written as
\begin{equation}
   \gamma_5H+H\gamma_5=2H^2f(H^2),
\label{eq:twoxtwo}
\end{equation}
where $f(H^2)$ is a regular function of $H^2$ and
$f(H^2)^\dagger=f(H^2)$. For simplicity, we assume that $f(x)$ is
monotonous and non-decreasing for~$x\geq0$. The simplest choice
$f(H^2)=1$ corresponds to the conventional Ginsparg-Wilson
relation~\cite{Ginsparg:1982bj}. The explicit form of~$H$ has been
analyzed for $f(H^2)=H^{2k}$ with a positive
integer~$k$~\cite{Fujikawa:2000my}. The following formal analyses are
valid for general $f(H^2)$, and one can recover the standard overlap
operator by setting $f(H^2)=1$ at any stage of our analyses. As an
important consequence of~eq.~(\ref{eq:twoxtwo}), one has
\begin{equation}
   \gamma_5H^2=(\gamma_5H+H\gamma_5)H-H(\gamma_5H+H\gamma_5)
   +H^2\gamma_5=H^2\gamma_5.
\label{eq:twoxthree}
\end{equation}
With this general Ginsparg-Wilson relation, one may introduce a
one-parameter family of lattice analogue of $\gamma_5$:
\begin{equation}
   \gamma_5^{(t)}={\gamma_5-tHf(H^2)\over\sqrt{1+t(t-2)H^2f^2(H^2)}}.
\label{eq:twoxfour}
\end{equation}
Note that $\gamma_5^{(0)}=\gamma_5$ and, when $f(H^2)=1$,
$\gamma_5^{(2)}=\hat\gamma_5$ corresponds to the conventional modified
chiral matrix~\cite{Niedermayer:1999bi}. Also the combination
\begin{equation}
   \Gamma_5=\gamma_5-Hf(H^2),
\label{eq:twoxfive}
\end{equation}
has a special role due to the property
\begin{equation}
   \Gamma_5H+H\Gamma_5=0.
\label{eq:twoxsix}
\end{equation}
Note that $\gamma_5^{(1)}=\Gamma_5/\sqrt{\Gamma_5^2}$. The ``conjugate''
of $\gamma_5^{(t)}$ is defined by
\begin{equation}
   \overline\gamma_5^{(t)}=\gamma_5\gamma_5^{(2-t)}\gamma_5,
\label{eq:twoxseven}
\end{equation}
and they satisfy the following relations:
\begin{eqnarray}
   &&\gamma_5^{(t)\dagger}=\gamma_5^{(t)},\qquad
   \overline\gamma_5^{(t)\dagger}=\overline\gamma_5^{(t)},\qquad
   (\gamma_5^{(t)})^2=(\overline\gamma_5^{(t)})^2=1,
\nonumber\\
   &&\overline\gamma_5^{(t)}D+D\gamma_5^{(t)}=0.
\label{eq:twoxeight}
\end{eqnarray}
In view of the last relation, we introduce the chiral projection
operators by
\begin{equation}
   P_\pm^{(t)}={1\over2}(1\pm\gamma_5^{(t)}),\qquad
   \overline P_\pm^{(t)}={1\over2}(1\pm\overline\gamma_5^{(t)}),
\label{eq:twoxnine}
\end{equation}
so that
\begin{equation}
   \overline P_\pm^{(t)}D=DP_\mp^{(t)}.
\label{eq:twoxten}
\end{equation}
Then the chirality may be defined by\footnote{For definiteness, we
consider the left-handed Weyl fermion.}
\begin{equation}
   P_-^{(t)}\psi=\psi,\qquad\overline\psi\overline P_+^{(t)}
=\overline\psi.
\label{eq:twoxeleven}
\end{equation}
The kinetic term is then consistently decomposed according to the
chirality.

\subsection{Generating functional and the measure term}

In the formulation of refs.~\cite{Luscher:1999du,Luscher:2000un}, the
expectation value of an operator~${\cal O}$ is given by
\begin{equation}
   \langle{\cal O}\rangle
   ={1\over{\cal Z}}\sum_M\int_M{\rm D}[U]\,e^{-S_G}
   {\cal N}_Me^{i\vartheta_M}\langle{\cal O}\rangle_{\rm F}^M,
\label{eq:twoxtwelve}
\end{equation}
where $M$ denotes the topological sector specified by the
admissibility,\footnote{Throughout this paper, we assume that the
representation of the gauge group, $R$, is unitary.}
\begin{equation}
   \|1-R[U(x,\mu,\nu)]\|<\epsilon,
   \qquad\hbox{for all plaquettes $(x,\mu,\nu)$,}
\label{eq:twoxthirteen}
\end{equation}
which guarantees the locality\footnote{The locality of the operator in
eq.~(\ref{eq:twoxtwo}) for general~$f(H^2)$ is not known. The locality
of free~$H$ for $f(H^2)=H^{2k}$ with a positive integer $k$ has been
established, and the topological properties of~$H$ for $f(H^2)=H^{2k}$
with a positive integer $k$ are known to be identical to those of the
standard overlap operator~\cite{Fujikawa:2000my}.} and the smoothness of
the Dirac operator~\cite{Hernandez:1999et,Neuberger:2000pz}. The
``topological weight'', ${\cal N}_Me^{i\vartheta_M}$, with which the
topological sectors are summed, is not fixed within this formulation. In
each topological sector~$M$, the average with respect to the fermion
fields~$\langle{\cal O}\rangle_{\rm F}^M$ is defined by the generating
functional of fermion Green's functions
\begin{equation}
   Z_{\rm F}[U,\eta,\overline\eta;t]
   =\int{\rm D}[\psi]{\rm D}[\overline\psi]\,e^{-S_{\rm F}},
\label{eq:twoxfourteen}
\end{equation}
where we have written the dependence on the parameter~$t$ in
eq.~(\ref{eq:twoxfour}) explicitly and
\begin{equation}
   S_{\rm F}=a^4\sum_x[\overline\psi(x)D\psi(x)-\overline\psi(x)\eta(x)
   -\overline\eta(x)\psi(x)].
\label{eq:twoxfifteen}
\end{equation}
The fermion integration variables are subject to the chirality
constraint~(\ref{eq:twoxeleven}).\footnote{We first analyze the case of
pure chiral gauge theory without Higgs couplings, and we shall later
analyze the case with Higgs couplings in Section~6.}

To define the fermion integration measure, we thus introduce certain
orthonormal vectors in the constrained space,
\begin{equation}
   P_-^{(t)}v_j(x)=v_j(x),\qquad
   \overline v_k\overline P_+^{(t)}(x)=\overline v_k(x),
\label{eq:twoxsixteen}
\end{equation}
and expand the fields as
\begin{equation}
   \psi(x)=\sum_jv_j(x)c_j,\qquad
   \overline\psi(x)=\sum_k\overline c_k\overline v_k(x).
\label{eq:twoxseventeen}
\end{equation}
Then the (ideal) integration measure is defined by
\begin{equation}
   {\rm D}[\psi]{\rm D}[\overline\psi]
   =\prod_j{\rm d}c_j\prod_k{\rm d}\overline c_k.
\label{eq:twoxeighteen}
\end{equation}
The condition~(\ref{eq:twoxsixteen}) shows that the basis vectors may
depend on the gauge field because the chiral projectors depend on it.
However, the condition~(\ref{eq:twoxsixteen}) alone does not specify the
fermion measure uniquely and there is an ambiguity of the phase. For two
different choices of basis, $\{v,\overline v\}$ and~$\{w,\overline w\}$,
we have
\begin{equation}
   Z_{\rm F}^{\{v,\overline v\}}[U,\eta,\overline\eta;t]
   =e^{i\theta[U;t]}
   Z_{\rm F}^{\{w,\overline w\}}[U,\eta,\overline\eta;t],
\label{eq:twoxnineteen}
\end{equation}
where the phase is given by the Jacobian for the change of basis
\begin{eqnarray}
   e^{i\theta[U;t]}&=&\det{\cal Q}\det\overline{\cal Q}
\nonumber\\
   &=&\exp(\Tr\ln{\cal Q}+\Tr\ln\overline{\cal Q}),
\label{eq:twoxtwenty}
\end{eqnarray}
where ${\cal Q}_{jk}=(w_j,v_k)$ and~$\overline{\cal Q}_{jk}=%
(\overline v_j^\dagger,\overline w_k^\dagger)$. Note that the phase
depends only on the gauge field. Under an infinitesimal variation of the
gauge field,
\begin{equation}
   \delta_\eta U(x,\mu)=a\eta_\mu(x)U(x,\mu),
\label{eq:twoxtwentyone}
\end{equation}
the variation of the phase is given by\footnote{This is derived from
\begin{eqnarray}
   i\theta[U;t]+i\delta_\eta\theta[U;t]
   &=&\Tr\ln(w_j+\delta_\eta w_j,v_k+\delta_\eta v_k)
   +\Tr\ln(\overline v_j^\dagger+\delta_\eta\overline v_j^\dagger,
   \overline w_k^\dagger+\delta_\eta\overline w_k^\dagger)
\nonumber\\
   &=&i\theta[U;t]-i{\mathfrak L}_\eta^{\{v,\overline v\}}[U;t]
   +i{\mathfrak L}_\eta^{\{w,\overline w\}}[U;t].
\label{eq:twoxtwentytwo}
\end{eqnarray}
}
\begin{equation}
   \delta_\eta\theta[U;t]
   =-{\mathfrak L}_\eta^{\{v,\overline v\}}[U;t]
   +{\mathfrak L}_\eta^{\{w,\overline w\}}[U;t],
\label{eq:twoxtwentythree}
\end{equation}
where the ``measure term'' is defined by
\begin{equation}
   {\mathfrak L}_\eta^{\{v,\overline v\}}[U;t]
   =i\sum_j(v_j,\delta_\eta v_j)
   +i\sum_k(\delta_\eta\overline v_k^\dagger,\overline v_k^\dagger).
\label{eq:twoxtwentyfour}
\end{equation}

To see the physical meaning of the measure term, let us temporarily
assume that there is no zero-modes of the Dirac operator. Then the
effective action without fermion sources is given by
\begin{equation}
   \langle1\rangle_{\rm F}=\det M,\qquad
   M_{kj}=a^4\sum_x\overline v_k(x)Dv_j(x),
\label{eq:twoxtwentyfive}
\end{equation}
and its variation under eq.~(\ref{eq:twoxtwentyone}) is given
by\footnote{We note that how the basis vectors~$v_j$ and~$\overline v_k$
change when the gauge field is varied cannot be determined by the
constraint~(\ref{eq:twoxsixteen}) alone. We first note the identity
$\delta_\eta v_j=P_-^{(t)}\delta_\eta v_j+%
(1-P_-^{(t)})\delta_\eta v_j$, and then note that the
constraint~(\ref{eq:twoxsixteen})
implies~$(1-P_-^{(t)})\delta_\eta v_j=\delta_\eta P_-^{(t)}v_j$ and
$\delta_\eta P_-^{(t)}$ is given by the construction of~$P_-^{(t)}$.
Consequently, the second term of the variation $\delta_\eta v_j$, 
$(1-P_-^{(t)})\delta_\eta v_j$, is fixed uniquely. However, the
component of~$\delta_\eta v_j$ residing in the constrained space,
$P_-^{(t)}\delta_\eta v_j$, is not determined by the constraint
(\ref{eq:twoxsixteen}). If one assumes that $v_j$ and~$\overline v_k$
change gauge covariantly under an infinitesimal gauge transformation,
$\delta_\eta v_j(x)=R(\omega(x))v_j(x)$
and~$\delta_\eta\overline v_k(x)=-\overline v_k(x)R(\omega(x))$, then
the effective action~(\ref{eq:twoxtwentyfive}) becomes invariant under
the gauge transformation. This assumption of the specific gauge
variation of (ideal) basis vectors, however, modifies the physical
contents of the theory in general, because it does not reproduce the
gauge anomaly. We will see such an example of basis vectors in Sec.~3.}
\begin{equation}
   \delta_\eta\ln\det M
   =\Tr(\delta_\eta DP_-^{(t)}D^{-1}\overline P_+^{(t)})
   -i{\mathfrak L}_\eta^{\{v,\overline v\}}.
\label{eq:twoxtwentysix}
\end{equation}
Since the measure term $-i{\mathfrak L}_\eta$ is pure imaginary, as its
definition~(\ref{eq:twoxtwenty}) implies, it contributes only to the
imaginary part of eq.~(\ref{eq:twoxtwentysix}):
\begin{eqnarray}
   &&{1\over2}[\Tr(\delta_\eta DP_-^{(t)}D^{-1}\overline P_+^{(t)})
   -\Tr(\delta_\eta DP_-^{(t)}D^{-1}\overline P_+^{(t)})^*]
\nonumber\\
   &&={1\over2}[
   \Tr(\delta_\eta DP_-^{(t)}D^{-1})
   -\Tr(\delta_\eta DP_+^{(2-t)}D^{-1})]
\nonumber\\
   &&=-{1\over4}\Tr\delta_\eta D(\gamma_5^{(t)}
   +\gamma_5^{(2-t)})D^{-1},
\label{eq:twoxtwentyseven}
\end{eqnarray}
where we have used $D^\dagger=\gamma_5D\gamma_5$
and~eq.~(\ref{eq:twoxseven}). The measure term is then chosen to improve
the imaginary part of the effective action (see, for example,
ref.~\cite{Suzuki:1999qw}).

The above general setup, which is closely related to the overlap
formulation in refs.~\cite{Narayanan:1993wx,Randjbar-Daemi:1995sq}, does
not specify the formulation uniquely, leaving the phase unspecified. The
crucial next task is to find an ideal basis in
eq.~(\ref{eq:twoxeighteen}) or the associated measure term for which the
gauge invariance (assuming the fermion multiplet is anomaly-free) and
the locality are ensured for finite lattice spacings. For $\U(1)$
theories, such a measure was constructed by
L\"uscher~\cite{Luscher:1999du}. For general non-abelian theories, the
measure has been constructed only in perturbation
theory~\cite{Luscher:2000zd,Suzuki:2000ii} but the existence of such an
ideal fermion measure in non-perturbative level has not been established
(except the case of electroweak $\SU(2)\times\U(1)$ on the infinite
volume lattice~\cite{Kikukawa:2001kd}). In this paper, we discuss the
effect of CP breaking in this formulation by simply assuming the
existence of such an ideal measure.\footnote{Note however that our
analysis is relevant for a manifestly gauge invariant perturbation
theory~\cite{Luscher:2000zd} based on this formulation.}

Our strategy to analyze the CP breaking is as follows: We first
determine the general structure of the fermion generating functional by
using a convenient auxiliary basis. Then using an argument of the change
of basis and a property of the measure term, we find the CP
transformation law of the generating functional.

\section{General structure of the generating functional}

\subsection{Generating functional with an auxiliary basis}

The relation~(\ref{eq:twoxnineteen}) shows that we may use any
basis~$\{w,\overline w\}$ as an intermediate tool in
analyzing~$Z_{\rm F}^{\{v,\overline v\}}$, if $e^{i\theta[U;t]}$
and~$Z_{\rm F}^{\{w,\overline w\}}$ are properly treated. A particularly
convenient basis is provided by the eigenfunctions of the hermitian
operator $D^\dagger D=H^2/a^2$:
\begin{equation}
   D^\dagger Du_j(x)
   ={1\over a^2}H^2u_j(x)={\lambda_j^2\over a^2}u_j(x),
\qquad
   \lambda_j\geq0,
\label{eq:threexone}
\end{equation}
and their appropriate projection
\begin{equation}
   w_j(x)=P_-^{(t)}u_j(x),\qquad(w_j,w_k)=\delta_{jk}.
\label{eq:threextwo}
\end{equation}
Note that $D^\dagger D$ and $P_-^{(t)}$ commute. From Appendix~B, where
the properties of these eigenfunctions are summarized, we see that the
following vectors have an appropriate chirality as~$w_j$:
\begin{eqnarray}
   &&\varphi_0^-(x),\qquad\lambda_j=0,
\nonumber\\
   &&\varphi_j^-(x),\qquad\lambda_j>0,\qquad\lambda_j\neq\Lambda,
\nonumber\\
   &&\cases{\Psi_+(x),& for $t>1$,\cr
            \Psi_-(x),& for $t<1$,\cr}\qquad\lambda_j=\Lambda,
\label{eq:threexthree}
\end{eqnarray}
where the number~$\Lambda$ is a solution of $\Lambda f(\Lambda^2)=1$.

As for the vectors $\overline w_k$, we may adopt the left eigenfunctions
of the hermitian operator~$DD^\dagger$.\footnote{Incidentally, the
Ginsparg-Wilson relation implies that
$DD^\dagger=\gamma_5H^2\gamma_5/a^2=H^2/a^2=D^\dagger D$.} For non-zero
eigenvalues, as is well-known, there is a one-to-one correspondence
between the eigenfunctions of $D^\dagger D$ and $DD^\dagger$:
\begin{equation}
   \overline w_j(x)={a\over\lambda_j}w_j^\dagger D^\dagger(x).
\label{eq:threexfour}
\end{equation}
This has the proper chirality as $\overline w_k$,
$\overline w_k\overline P_+^{(t)}=\overline w_k$. The zero-modes
of~$DD^\dagger$ cannot be expressed in this way and we may use instead
\begin{equation}
   \varphi_0^{+\dagger}(x).
\label{eq:threexfive}
\end{equation}
Then eqs.~(\ref{eq:threexfour}) and~(\ref{eq:threexfive}) span a
complete set in the constrained
space~$\overline w_k\overline P_+^{(t)}=\overline w_k$.

Once having specified basis vectors $\{w,\overline w\}$, it is
straightforward to perform the integration
in~$Z_{\rm F}^{\{w,\overline w\}}$. After some calculation, we have
\begin{eqnarray}
   &&Z_{\rm F}^{\{w,\overline w\}}[U,\eta,\overline\eta;t]
   =\left({\Lambda\over a}\right)^N
   \prod^{n_-}\left[a^4\sum_x\overline\eta(x)\varphi_0^-(x)\right]
   \prod^{n_+}\left[a^4\sum_x\varphi_0^{+\dagger}(x)\eta(x)\right]
   \prod_{\lambda_j>0\atop\lambda_j\neq\Lambda}
   \left({\lambda_j\over a}\right)
\nonumber\\
   &&\qquad\qquad\qquad\qquad
   \times
   \exp\left[a^8\sum_{x,y}\overline\eta(x)G^{(t)}(x,y)
   \eta(y)\right],
\label{eq:threexsix}
\end{eqnarray}
up to the over-all sign $\pm1$ which depends on the ordering in the 
measure $\prod_j\rmd c_j\prod_k\rmd\overline c_k$.\footnote{This sign
factor can be absorbed into the phase~$\theta[U;t]$ without loss of
generality.} In this expression, the Green's function has been defined
by
\begin{equation}
   DG^{(t)}(x,y)=\overline P_+^{(t)}(x,y)
   -\sum^{n_+}\varphi_0^+(x)\varphi_0^{+\dagger}(y),
\label{eq:threexseven}
\end{equation}
or more explicitly,
\begin{equation}
   G^{(t)}(x,y)=\sum_{\lambda_j>0}{a^2\over\lambda_j^2}\,
   \varphi_j^-(x)\varphi_j^{-\dagger}D^\dagger(y).
\label{eq:threexeight}
\end{equation}
The number of zero-modes $\varphi_0^-$ ($\varphi_0^+$) has been denoted
by $n_-$ ($n_+$) and
\begin{equation}
   N=\cases{N_+,&for $t>1$,\cr
            N_-,&for $t<1$,\cr}
\label{eq:threexnine}
\end{equation}
where $N_+$ and $N_-$ stand for the numbers of eigenfunctions $\Psi_+$
and~$\Psi_-$, respectively (see Appendix~B). Since eigenvalues
$\lambda_j$ are gauge invariant and eigenfunctions $\varphi_j$ can be
chosen to be gauge covariant, the above $Z_{\rm F}^{\{w,\overline w\}}$
is manifestly gauge invariant for gauge covariant external sources.
However, this~$Z_{\rm F}^{\{w,\overline w\}}$ as it stands cannot be
interpreted as the generating functional for the Weyl fermion, as we
will explain shortly (rather it is regarded as representing a half of
the Dirac fermion).

\subsection{Measure term for the auxiliary basis}

Let us consider the measure term~(\ref{eq:twoxtwentyfour}) for the
auxiliary basis. Namely,
\begin{equation}
   {\mathfrak L}_\eta^{\{w,\overline w\}}[U;t]
   =i\sum_j(w_j,\delta_\eta w_j)
   +i\sum_k(\delta_\eta\overline w_k^\dagger,\overline w_k^\dagger).
\label{eq:threexten}
\end{equation}
Noting $H^2w_j=\lambda_j^2w_j$ and thus
\begin{equation}
   \delta_\eta\lambda_j^2=(w_j,\delta_\eta H^2w_j),
\label{eq:threexeleven}
\end{equation}
we have from eq.~(\ref{eq:threexfour}),
\begin{equation}
   (\delta_\eta\overline w_j^\dagger,\overline w_j^\dagger)
   =-{1\over2\lambda_j^2}
   (w_j,[H,\delta_\eta H]w_j)+(\delta_\eta w_j, w_j).
\label{eq:threextwelve}
\end{equation}
Taking the contribution of zero-modes into account, we thus have
\begin{equation}
   {\mathfrak L}_\eta^{\{w,\overline w\}}[U;t]
   =i\sum^{n_-}(\varphi_0^-,\delta_\eta\varphi_0^-)
   +i\sum^{n_+}(\delta_\eta\varphi_0^+,\varphi_0^+)
   -{i\over2}
   \sum_{\lambda_j>0}(\varphi_j^-,[H^{-1},\delta_\eta H]\varphi_j^-).
\label{eq:threexthirteen}
\end{equation}
The last term may be written in a basis independent way
\begin{eqnarray}
   -{i\over2}\Tr'[H^{-1},\delta_\eta H]P_-^{(t)}
   &=&{i\over4}\Tr'(H^{-1}\delta_\eta H\gamma_5^{(t)}
   -\delta_\eta HH^{-1}\gamma_5^{(t)})
\nonumber\\
   &=&{i\over4}\Tr'\delta_\eta H
   (\gamma_5^{(t)}+\gamma_5^{(2-t)})H^{-1},
\label{eq:threexfourteen}
\end{eqnarray}
where $\Tr'$ denotes the trace over the subspace of non-zero modes of
the hermitian operator~$H$.\footnote{One thus has to be careful whether
the operator concerned preserves this subspace when using the cyclic
property of the trace.} In deriving the last line, we have used the
relation $H^{-1}\gamma_5^{(t)}=-\gamma_5^{(2-t)}H^{-1}$ being valid in
this subspace. Noting $H^{-1}\gamma_5\gamma_5\delta_\eta H=%
D^{-1}\delta_\eta D$, we have
\begin{equation}
   {\mathfrak L}_\eta^{\{w,\overline w\}}[U;t]
   =i\sum^{n_-}(\varphi_0^-,\delta_\eta\varphi_0^-)
   +i\sum^{n_+}(\delta_\eta\varphi_0^+,\varphi_0^+)
   +{i\over4}\Tr'\delta_\eta D(\gamma_5^{(t)}+\gamma_5^{(2-t)})D^{-1}.
\label{eq:threexfifteen}
\end{equation}

This expression shows that the auxiliary basis $w_j$ and~$\overline w_k$
cannot be a physically sensible one (namely, we cannot take
$\{v,\overline v\}=\{w,\overline w\}$) because the measure term is
non-local, containing the propagator~$D^{-1}$. In fact, we see that
$i{\mathfrak L}_\eta^{\{w,\overline w\}}$ is identical to (a variation
of) the main part of the imaginary part of the fermion effective
action~(\ref{eq:twoxtwentyseven}), when there is no zero-modes.
Consequently, this basis when identified as
$\{v,\overline v\}=\{w,\overline w\}$ modifies the physical contents of
the theory, eliminating the imaginary part. This explains why the
generating functional with this basis is gauge invariant, even if the
fermion multiplet is not anomaly-free. Nevertheless, this basis is
convenient as an intermediate tool as one can work out all the
quantities.

\subsection{$Z_{\rm F}^{\{v,\overline v\}}$}

{}From eqs.~(\ref{eq:twoxnineteen}) and~(\ref{eq:threexsix}), the
general structure of the fermion generating functional is given by
(omitting the superscript ${\{v,\overline v\}}$)
\begin{eqnarray}
   &&Z_{\rm F}[U,\eta,\overline\eta;t]
   =e^{i\theta[U;t]}\left({\Lambda\over a}\right)^N
   \prod^{n_-}\left[a^4\sum_x\overline\eta(x)\varphi_0^-(x)\right]
   \prod^{n_+}\left[a^4\sum_x\varphi_0^{+\dagger}(x)\eta(x)\right]
   \prod_{\lambda_j>0\atop\lambda_j\neq\Lambda}
   \left({\lambda_j\over a}\right)
\nonumber\\
   &&\qquad\qquad\qquad\qquad
   \times
   \exp\left[a^8\sum_{x,y}\overline\eta(x)G^{(t)}(x,y)
   \eta(y)\right],
\label{eq:threexsixteen}
\end{eqnarray}
where the variation of the phase $\theta[U;t]$ is given by
eq.~(\ref{eq:twoxtwentythree}) and the measure term for the auxiliary
basis is given by eq.~(\ref{eq:threexfifteen}). We see that the vital
characterization as the {\it chiral\/} theory is contained in the
phase~$\theta[U;t]$ which may be computed, only after finding the ideal
basis ${\{v,\overline v\}}$ (or the associated measure). For our
discussion of CP breaking, however, only a certain property of the
measure term ${\mathfrak L}_\eta^{\{v,\overline v\}}[U;t]$ will turn out
to be sufficient.

\section{CP transformed generating functional}

\subsection{Generating functional and CP transformation}

We first note
\begin{eqnarray}
   &&\|1-R[U^{\rm CP}(x,i,j)]\|
   =\|1-R[U(\bar x-a\hat i-a\hat j,i,j)]\|,
\nonumber\\
   &&\|1-R[U^{\rm CP}(x,i,4)]\|=\|1-R[U(\bar x-a\hat i,i,4)]\|,
\label{eq:fourxone}
\end{eqnarray}
according to the CP transformation law of the plaquette variables in
Appendix~A. Thus, if $U$ is an admissible configuration, so is
$U^{\rm CP}$; CP preserves the admissibility. Note however that $U$
and~$U^{\rm CP}$ may belong to different topological sectors in general.
In fact, the index~$n_+-n_-$~\cite{Hasenfratz:1998ri} is opposite
for~$U$ and for~$U^{\rm CP}$ (see Appendix~B):
\begin{equation}
   \Tr\Gamma_5(U)=n_+-n_-=-\Tr\Gamma_5(U^{\rm CP}),
\label{eq:fourxtwo}
\end{equation}
from $\Gamma_5(U^{\rm CP})=-W^{-1}(\gamma_5\Gamma_5(U)\gamma_5)^TW$.%
\footnote{We have
\begin{equation}
   H(U^{\rm CP})=-W^{-1}(\gamma_5H(U)\gamma_5)^TW,
\label{eq:fourxthree}
\end{equation}
and
\begin{eqnarray}
   &&\gamma_5^{(t)}(U^{\rm CP})
   =-W^{-1}\overline\gamma_5^{(2-t)}(U)^TW,\qquad
   \overline\gamma_5^{(t)}(U^{\rm CP})=-W^{-1}\gamma_5^{(2-t)}(U)^TW,
\nonumber\\
   &&P_\pm^{(t)}(U^{\rm CP})=W^{-1}\overline P_\mp^{(2-t)}(U)^TW,\qquad
   \overline P_\pm^{(t)}(U^{\rm CP})=W^{-1}P_\mp^{(2-t)}(U)^TW,
\label{eq:fourxfour}
\end{eqnarray}
from the CP transformation law in Appendix~A. Throughout this paper, the
transpose operation and the complex conjugation of an operator are
defined with respect to the corresponding kernel in position space.
Strictly speaking, the coordinates~$x$ in these expressions are replaced
by $\bar x=(-x_i,x_4)$ under CP. We can forgo writing this explicitly,
because our final expressions always involve an integration over~$x$ and
$\sum_x=\sum_{\bar x}$.}

Now, let us consider the CP transformed generating functional%
\footnote{It is possible to write this formula as
\begin{eqnarray}
   &&Z_{\rm F}[U^{\rm CP},-W^{-1}\overline\eta^T,\eta^TW;t]
\label{eq:fourxfive}
\\
   &&=\int{\rm D}[\psi^{\rm CP}]{\rm D}[\overline\psi^{\rm CP}]\,
   \exp\biggl\{-a^4\sum_x[
   \overline\psi^{\rm CP}(x)D(U^{\rm CP})\psi^{\rm CP}(x)
   +\overline\psi^{\rm CP}(x)W^{-1}\overline\eta^T(x)
   -\eta^T(x)W\psi^{\rm CP}(x)]\biggr\}.
\nonumber
\end{eqnarray}
One thus sees that there are two possible sources of CP violation: An
explicit breaking in the action and an anomalous breaking in the path
integral measure.}
\begin{equation}
   Z_{\rm F}[U^{\rm CP},-W^{-1}\overline\eta^T,\eta^TW;t]
   =\int{\rm D}[\psi]{\rm D}[\overline\psi]\,e^{-S_{\rm F}},
\label{eq:fourxsix}
\end{equation}
where
\begin{equation}
   S_{\rm F}=a^4\sum_x[\overline\psi(x)D(U^{\rm CP})\psi(x)
   +\overline\psi(x)W^{-1}\overline\eta^T(x)
   -\eta^T(x)W\psi(x)],
\label{eq:fourxseven}
\end{equation}
and~${\rm D}[\psi]{\rm D}[\overline\psi]=%
\prod_j\rmd c_j\prod_k\rmd\overline c_k$. Here the ideal basis vectors
in $\psi(x)=\sum_jv_j(x)c_j$ and~$\overline\psi(x)=%
\sum_k\overline c_k\overline v_k(x)$ are defined through the
constraints:
\begin{equation}
   P_-^{(t)}(U^{\rm CP})v_j(x)=v_j(x),
   \qquad\overline v_k(x)\overline P_+^{(t)}(U^{\rm CP})
   =\overline v_k(x).
\label{eq:fourxeight}
\end{equation}
The generating functional~(\ref{eq:fourxsix}) can then be written as
\begin{eqnarray}
   &&Z_{\rm F}[U^{\rm CP},-W^{-1}\overline\eta^T,\eta^TW;t]
\label{eq:fourxnine}
\\
   &&=\int\prod_j\rmd c_j\prod_k\rmd\overline c_k\,
   \exp\biggl\{-a^4\sum_x[\overline\psi'(x)D(U)\psi'(x)
   -\overline\eta(x)\psi'(x)-\overline\psi'(x)\eta(x)]\biggr\},
\nonumber
\end{eqnarray}
where
\begin{eqnarray}
   &&\psi'(x)=[\overline\psi(\bar x)W^{-1}]^T
   =\sum_k\overline c_k[\overline v_k(\bar x)W^{-1}]^T,
\nonumber\\
   &&\overline\psi'(x)=[-W\psi(\bar x)]^T
   =\sum_j[-Wv_j(\bar x)]^Tc_j.
\label{eq:fourxten}
\end{eqnarray}
Since the basis vectors in eq.~(\ref{eq:fourxten}),
\begin{equation}
   v_k'=(\overline v_kW^{-1})^T,\qquad
   \overline v_j'=(-Wv_j)^T,
\label{eq:fourxeleven}
\end{equation}
satisfy the constraints
\begin{equation}
   P_-^{(2-t)}(U)v_k'=v_k',\qquad
   \overline v_j'\overline P_+^{(2-t)}(U)
   =\overline v_j',
\label{eq:fourxtwelve}
\end{equation}
a comparison of eq.~(\ref{eq:fourxnine}) with the original generating
functional~(\ref{eq:twoxfourteen}) shows
\begin{equation}
   Z_{\rm F}[U^{\rm CP},-W^{-1}\overline\eta^T,\eta^TW;t]
   =Z_{\rm F}[U,\eta,\overline\eta;2-t].
\label{eq:fourxthirteen}
\end{equation}
Thus the sole effect of the CP transformation is given by the change of
the parameter, $t\to2-t$. Instead of repeating the calculation in Sec.~3
for~$U^{\rm CP}$, it is thus enough to examine the effect of~$t\to2-t$
in eq.~(\ref{eq:threexsixteen}).\footnote{The dimensionality of
fermionic spaces before and after CP transformation is however
different,
\begin{equation}
   \Tr[\overline P_+^{(2-t)}(U)+P_-^{(2-t)}(U)]
   -\Tr[\overline P_+^{(t)}(U)+P_-^{(t)}(U)]
   =2{t-1\over|t-1|}(n_+-n_-),
\label{eq:fourxfourteen}
\end{equation}
namely, the dimensionality jumps at $t=1$ in the presence of
topologically non-trivial gauge field.}

In eq.~(\ref{eq:threexsixteen}), we note that the zero-modes
$\varphi_0^-$, $\varphi_0^+$ and the eigenvalues~$\lambda_j$ are
independent of~$t$ (see Appendix~B). The change $t\to2-t$, however,
causes the exchange of $\Psi_+$ and~$\Psi_-$ as shown in
eq.~(\ref{eq:threexthree}) and thus $N\to\overline N$, where
\begin{equation}
   \overline N=\cases{N_-&for $t>1$,\cr
            N_+&for $t<1$.\cr}
\label{eq:fourxfifteen}
\end{equation}
Thus, we immediately have
\begin{eqnarray}
   &&Z_{\rm F}[U^{\rm CP},-W^{-1}\overline\eta^T,\eta^TW;t]
\nonumber\\
   &&=e^{i\theta[U;2-t]}\left({\Lambda\over a}\right)^{\overline N}
   \prod^{n_-}\left[a^4\sum_x\overline\eta(x)\varphi_0^-(x)\right]
   \prod^{n_+}\left[a^4\sum_x\varphi_0^{+\dagger}(x)\eta(x)\right]
   \prod_{\lambda_j>0\atop\lambda_j\neq\Lambda}
   \left({\lambda_j\over a}\right)
\nonumber\\
   &&\qquad\qquad\qquad\qquad
   \times
   \exp\left[a^8\sum_{x,y}\overline\eta(x)G^{(2-t)}(x,y)
   \eta(y)\right],
\label{eq:fourxsixteen}
\end{eqnarray}
for the generating functional.

As for the phase factor~$e^{\theta[U;t]}$, its variation is given by
eq.~(\ref{eq:twoxtwentythree}). The measure terms for the auxiliary
basis~${\mathfrak L}_\eta^{\{w,\overline w\}}$ is invariant
under~$t\to2-t$:
\begin{equation}
   {\mathfrak L}_\eta^{\{w,\overline w\}}[U;2-t]
   ={\mathfrak L}_\eta^{\{w,\overline w\}}[U;t],
\label{eq:fourxseventeen}
\end{equation}
as can readily be seen from eq.~(\ref{eq:threexfifteen}). Moreover, in
the next section, we will show that it is always possible to choose the
ideal basis vectors such that
\begin{equation}
   {\mathfrak L}_\eta^{\{v,\overline v\}}[U;2-t]
   ={\mathfrak L}_\eta^{\{v,\overline v\}}[U;t].
\label{eq:fourxeighteen}
\end{equation}
As a result, we have
\begin{equation}
   \delta_\eta(\theta[U;2-t]-\theta[U;t])=0,
\label{eq:fourxnineteen}
\end{equation}
and the difference, $\theta[U;2-t]-\theta[U;t]$, if it exists, is a
{\it constant}:
\begin{equation}
   \theta[U;2-t]=\theta[U;t]+\theta_M,
\label{eq:fourxtwenty}
\end{equation}
where the constant~$\theta_M$ is assigned for each topological
sector~$M$, $U\in M$.

{}From the above analysis, we have
\begin{eqnarray}
   &&Z_{\rm F}[U^{\rm CP},-W^{-1}\overline\eta^T,\eta^TW;t]
\nonumber\\
   &&=e^{i\theta_M}\left({\Lambda\over a}\right)^{\overline N-N}
   {\exp\left[a^8\sum_{x,y}\overline\eta(x)G^{(2-t)}(x,y)
   \eta(y)\right]\over
   \exp\left[a^8\sum_{x,y}\overline\eta(x)G^{(t)}(x,y)
   \eta(y)\right]}Z_{\rm F}[U,\eta,\overline\eta;t].
\label{eq:fourxtwentyone}
\end{eqnarray}
In particular, in the vacuum sector which contains the trivial
configuration $U_0(x,\mu)=1$, $U_0^{\rm CP}=U_0$ and thus one has
$\theta_M=0$ (recall that the phase depends only on the gauge field).

{}From eq.~(\ref{eq:fourxtwentyone}), we see that the CP breaking in
this formulation appears in three places: (I)~Difference in the overall
constant phase~$\theta_M$. (II)~Difference in the overall
coefficient~$(\Lambda/a)^{\overline N-N}$ (III)~Difference in the
propagator, $G^{(t)}$ and $G^{(2-t)}$. We discuss their implications in
this order:

(I)~The {\it constant\/} phase~$\theta_M$ may be absorbed into a
redefinition of the phase factor $\vartheta_M$ in
eq.~(\ref{eq:twoxtwelve}) as
\begin{equation}
   \vartheta_M\to\vartheta_M+{1\over2}\theta_M,
   \qquad\vartheta_{M^{\rm CP}}\to
   \vartheta_{M^{\rm CP}}-{1\over2}\theta_M.
\label{eq:fourxtwentytwo}
\end{equation}
Then the overall phases in $Z_{\rm F}[U]$ and $Z_{\rm F}[U^{\rm CP}]$
become identical (no ``CP anomaly'' from the path integral measure) and
the discussion of CP violation is reduced to how one should choose the
``topological phase'' $\vartheta_M$; this is a problem analogous to the
strong CP problem.

(II)~The breaking~$(\Lambda/a)^{\overline N-N}$ can also be absorbed
into the topological weight~${\cal N}_M$ in~eq.~(\ref{eq:twoxtwelve}).
Namely, we may redefine
\begin{equation}
   {\cal N}_M
   \to {\cal N}_M
   \left({\Lambda\over a}\right)^{(-N+\overline N)/2},\qquad
   {\cal N}_{M^{\rm CP}}\to {\cal N}_{M^{\rm CP}}
   \left({\Lambda\over a}\right)^{(N-\overline N)/2}.
\label{eq:fourxtwentythree}
\end{equation}
This redefinition is consistent because the roles of $N$ and
$\overline N$ are exchanged under $U\leftrightarrow U^{\rm CP}$. Note
that
\begin{equation}
   -N+\overline N=\cases{n_+-n_-,&for $t>1$,\cr
                         n_--n_+,&for $t<1$,\cr}
\label{eq:fourxtwentyfour}
\end{equation}
due to the chirality sum rule~\cite{Chiu:1998bh,Fujikawa:1999ku}. (The
index~$n_+-n_-$ does not depend on~$f(H^2)$, see Appendix~B.) The
simplest CP invariant choice is then ${\cal N}_M=1$ for all topological
sectors. However, whether this simplest choice is consistent with other
physical requirements, such as the cluster decomposition, is another
question which we do not discuss in this paper. Interestingly, this
simplest CP invariant choice is also suggested~\cite{Suzuki:2000ku} by a
matching with the ``Majorana formulation''.

(III)~It seems impossible to remedy the breaking $G^{(2-t)}\neq G^{(t)}$
in the propagator. Note that the propagator is independent of the choice
of the basis vectors or the path integral measure. For the symmetric
choice~$t=1$, $\gamma_5^{(1)}=\Gamma_5/\sqrt{\Gamma_5^2}$ is plagued
with the singularity due to zero-modes of $\Gamma_5$, whose inevitable
presence is proven under rather mild assumptions~\cite{Fujikawa:2002is}.
However, observe that the CP breaking for $t\neq1$ is quite modest. For
example, when there are no zero-modes,
\begin{equation}
   G^{(2-t)}
   =P_-^{(2-t)}{1\over D}\overline P_+^{(2-t)}
   =G^{(t)}
   +{a(1-t)f(H^2)\over\sqrt{1+t(t-2)H^2f^2(H^2)}}\gamma_5,
\label{eq:fourxtwentyfive}
\end{equation}
thus the breaking term is local. In particular, for the conventional 
choice, $t=2$ and $f(H^2)=1$,
\begin{equation}
   G^{(2-t)}(x,y)
   =G^{(t)}(x,y)-a\gamma_5{1\over a^4}\delta_{x,y},
\label{eq:fourxtwentysix}
\end{equation}
and the breaking appears as an (ultra-local) contact term, as we have
noted in Introduction. It is thus expected that this breaking is safely
removed in a suitable continuum limit in the case of pure chiral gauge
theory. However, there appear additional complications when the Yukawa
coupling is included and the Higgs field acquires the expectation
value; this issue will be discussed in Sec.~6.

In summary, the inherent CP violation in this framework emerges only in
the fermion propagator which is connected to external sources. This
implies that diagrams with external fermion lines or with a fermion
composite operator would behave differently from the naively expected
one under CP, but the vacuum polarization, for example, respects CP. As
for other possible sources of CP violation in relative topological
weight factors, the same problem appears in continuum theory also and it
is not particular to the present formulation of lattice chiral gauge
theory.

\section{Property of the ideal measure term}

\subsection{Reconstruction theorem%
\protect\footnote{This sub-section gives a brief sketch of the result of
extensive analyses. Those who are interested in further details are
asked to refer to refs.~\cite{Luscher:1999du,Luscher:2000un}. For our
analysis of CP symmetry, the relation in eq.~(\ref{eq:fivexten}) or 
eq.~(\ref{eq:fivexfifteen}) is essential.}}

In our formulation, the basis vectors~$\overline v_k(x)$
for~$\overline\psi(x)$ may also depend on the gauge field. Thus, to
accommodate this case, we need to slightly generalize the reconstruction
theorem~\cite{Luscher:1999du,Luscher:2000un} which represents precise
conditions for the ``ideal'' measure term.

The measure term~${\mathfrak L}_\eta$ can be interpreted as the $\U(1)$
connection associated to a fiber bundle defined over the space of
admissible configurations~\cite{Neuberger:1999xn,Luscher:1999du}. The
$\U(1)$ connection is characterized by its ``curvature''\footnote{Here
the variations $\eta$ and~$\zeta$ are assumed to be independent of the
gauge field.}
\begin{eqnarray}
   \delta_\eta{\mathfrak L}_\zeta-\delta_\zeta{\mathfrak L}_\eta
   +a{\mathfrak L}_{[\eta,\zeta]}
   &=&i\Tr P_-^{(t)}[\delta_\eta P_-^{(t)},\delta_\zeta P_-^{(t)}]
   +i\Tr\overline P_+^{(t)}
   [\delta_\zeta\overline P_+^{(t)},\delta_\eta\overline P_+^{(t)}]
\label{eq:fivexone}
\\
   &=&-{i\over8}\Tr\gamma_5^{(t)}
   [\delta_\eta\gamma_5^{(t)},\delta_\zeta\gamma_5^{(t)}]
   -{i\over8}\Tr\gamma_5^{(2-t)}
   [\delta_\eta\gamma_5^{(2-t)},\delta_\zeta\gamma_5^{(2-t)}],
\nonumber
\end{eqnarray}
and, by the ``Wilson line'' along a curve $U_\tau(x,\mu)$
($0\leq\tau\leq1$),
\begin{equation}
   W=\exp\biggl(i\int_0^1\rmd\tau\,{\mathfrak L}_\eta\biggr),\quad
   a\eta_\mu(x)=\partial_\tau U_\tau(x,\mu)U_\tau(x,\mu)^{-1}.
\label{eq:fivextwo}
\end{equation}
Assuming that the fermion measure is smoothly defined over the space of
admissible configurations, the Wilson line along a {\it loop}, for which
$U_1=U_0$, is given by (see ref.~\cite{Luscher:2000un})
\begin{eqnarray}
   W&=&\det(1-P_0^{(t)}+P_0^{(t)}Q_1^{(t)})
   \det\nolimits^{-1}(1-\overline P_0^{(t)}
   +\overline P_0^{(t)}\gamma_5Q_1^{(2-t)}\gamma_5)
\nonumber\\
   &=&\det(1-P_0^{(t)}+P_0^{(t)}Q_1^{(t)})
   \det\nolimits(1-P_0^{(2-t)}+P_0^{(2-t)}Q_1^{(2-t)}),
\label{eq:fivexthree}
\end{eqnarray}
where $P_\tau^{(t)}=P_-^{(t)}|_{U=U_\tau}$
and~$\overline P_\tau^{(t)}=\overline P_+^{(t)}|_{U=U_\tau}$. The
unitary operator~$Q_\tau^{(t)}$ in this expression is defined by
\begin{equation}
   \partial_\tau Q_\tau^{(t)}
   ={1\over4}[\partial_\tau\gamma_\tau^{(t)},\gamma_\tau^{(t)}]
   Q_\tau^{(t)},\qquad Q_0^{(t)}=1,
\label{eq:fivexfour}
\end{equation}
where $\gamma_\tau^{(t)}=\gamma_5^{(t)}|_{U=U_\tau}$. From this, we have
$\gamma_\tau^{(t)}=Q_\tau^{(t)}\gamma_0^{(t)}Q_\tau^{(t)\dagger}$ and
thus
\begin{equation}
   P_\tau^{(t)}=Q_\tau^{(t)}P_0^{(t)}Q_\tau^{(t)\dagger},\qquad
   \overline P_\tau^{(t)}
   =\gamma_5Q_\tau^{(2-t)}\gamma_5\overline P_0^{(t)}
   (\gamma_5Q_\tau^{(2-t)}\gamma_5)^\dagger.
\label{eq:fivexfive}
\end{equation}
These relations have been used in deriving the first line of
eq.~(\ref{eq:fivexthree}). From the first line to the second line in
eq.~(\ref{eq:fivexthree}), we have noted $\det Q_\tau^{(t)}=1$.

The gauge variation of the expectation value, on the other hand, is
given by
\begin{eqnarray}
   &&\delta_\eta\langle{\cal O}\rangle_{\rm F}
   =\langle\delta_\eta{\cal O}\rangle_{\rm F}
   +ia^4\sum_x\omega^a(x)[{\cal A}^a(x)-(\nabla_\mu^*j_\mu)^a(x)]
   \langle{\cal O}\rangle_{\rm F},
\nonumber\\
   &&{\cal A}^a(x)=-{i\over2}\tr R(T^a)
   (\gamma_5^{(t)}+\gamma_5^{(2-t)})(x,x),
\label{eq:fivexsix}
\end{eqnarray}
by setting $\eta_\mu(x)=-\nabla_\mu\omega(x)$
and~${\mathfrak L}_\eta=a^4\sum_x\eta_\mu^a(x)j_\mu^a(x)$.%
\footnote{Covariant derivatives in these expressions are defined by
\begin{eqnarray}
   &&\nabla_\mu\omega(x)
   ={1\over a}[U(x,\mu)\omega(x+a\hat\mu)U(x,\mu)^{-1}-\omega(x)],
\nonumber\\
   &&\nabla_\mu^*j_\mu(x)
   ={1\over a}[j_\mu(x)-U(x-a\hat\mu,\mu)^{-1}j_\mu(x-a\hat\mu)
   U(x-a\hat\mu,\mu)].
\label{eq:fivexseven}
\end{eqnarray}
}
So, for the gauge invariant measure, the measure term should satisfy the
``anomalous conservation law'':
\begin{equation}
   (\nabla_\mu^*j_\mu)^a(x)={\cal A}^a(x).
\label{eq:fivexeight}
\end{equation}

Now, we have observed that for a smooth gauge invariant fermion measure,
the associated measure term satisfies the conditions,
eqs.~(\ref{eq:fivexthree}) and~(\ref{eq:fivexeight})
[eq.~(\ref{eq:fivexone}) can be derived from eq.~(\ref{eq:fivexthree})].
Conversely, these two conditions are in fact sufficient. Namely, if one
has a certain current~$j_\mu^a(x)$ such that eq.~(\ref{eq:fivexeight})
is fulfilled and the combination
${\mathfrak L}_\eta=a^4\sum_x\eta_\mu^a(x)j_\mu^a(x)$ satisfies
eq.~(\ref{eq:fivexthree}) for any loop, then there exists a smooth gauge
invariant fermion measure. This is the reconstruction theorem and the
proof is given by an explicit construction: We may set the fermionic
basis vectors for the gauge field $U$ as
\begin{equation}
   v_j=\cases{Q_1^{(t)}u_1W^{-1},&for $j=1$,\cr
              Q_1^{(t)}u_j,&otherwise,\cr}\qquad
   \overline v_k=\overline u_k(\gamma_5Q_1^{(2-t)}\gamma_5)^\dagger,
\label{eq:fivexnine}
\end{equation}
where $u_j$ and~$\overline u_k$ are fixed bases defined for the
reference configuration~$U_0$. The vectors are then transported by the
operators~$Q_\tau^{(t)}$ and~$Q_\tau^{(2-t)}$ along a certain curve
connecting $U_0$ and~$U=U_1$. The Wilson line~$W$ is also defined along
this curve. The above basis depends on the curve chosen, but the
associated measure does not, because of~eq.~(\ref{eq:fivexthree}). Thus
the measure is smooth. The measure term for the
basis~(\ref{eq:fivexnine}) is given by~${\mathfrak L}_\eta$. Thus the
anomalous conservation law is also fulfilled.

\subsection{Invariance under $t\to2-t$ and the CP property of the
measure term}

In the last subsection, we have observed that the requirements for the
ideal measure term are given by eqs.~(\ref{eq:fivexthree})
and~(\ref{eq:fivexeight}).\footnote{Another important requirement is
that the measure term must be local, to be consistent with the locality
of the theory.} The remarkable fact is that these conditions are
invariant under $t\to2-t$. Thus, if we have an ideal measure
term~${\mathfrak L}_\eta[U;t]$ which works for $U$ with respect
to~$\gamma_5^{(t)}$, then we may use the {\it same\/} measure term
for~$U$ with respect to~$\gamma_5^{(2-t)}$. Therefore, we may set
without loss of generality
\begin{equation}
   {\mathfrak L}_\eta[U;2-t]={\mathfrak L}_\eta[U;t].
\label{eq:fivexten}
\end{equation}
This is the equality~(\ref{eq:fourxeighteen}) we have used in the
previous section.

This equality can be interpreted in a more physically transparent
language; this is equivalent to the CP invariance of the measure term.
To see this, let us recall that basis vectors for~$U^{\rm CP}$ with
respect to~$\gamma_5^{(t)}$ [which is specified by
eq.~(\ref{eq:fourxeight})] and basis vectors for~$U$ with respect
to~$\gamma_5^{(2-t)}$ [which is specified by eq.~(\ref{eq:fourxtwelve})]
can be related as eq.~(\ref{eq:fourxeleven}). This particular choice of
bases, which may always be made, leads to
\begin{equation}
   {\mathfrak L}_\eta[U^{\rm CP};t]={\mathfrak L}_\eta[U;2-t],
\label{eq:fivexeleven}
\end{equation}
and thus eq.~(\ref{eq:fivexten}) implies
\begin{equation}
   {\mathfrak L}_\eta[U^{\rm CP};t]={\mathfrak L}_\eta[U;t].
\label{eq:fivextwelve}
\end{equation}
In fact, it is physically natural to take basis vectors such that the
relation~(\ref{eq:fivextwelve}) holds. In the continuum theory, the
action of the Weyl fermion is CP invariant and the imaginary part of the
effective action is too (it is independent of the regularization chosen
and is given by the so-called
$\eta$-invariant~\cite{Alvarez-Gaume:1985di}). This property is shared
with our lattice transcription~(\ref{eq:twoxtwentyseven}), as one can
verify from CP transformation law of various operators.\footnote{One
should however be careful about the meaning of the
variation~$\delta_\eta$.
Under $U(x,\mu)\to U(x,\mu)+\delta_\eta U(x,\mu)$, the CP transformed
configuration changes as $U^{\rm CP}(x,\mu)\to U^{\rm CP}(x,\mu)%
+\delta_\eta U^{\rm CP}(x,\mu)$. With this understanding, defining
\begin{equation}
   \delta_\eta U^{\rm CP}(x,\mu)
   =a\eta_\mu^{\rm CP}(x)U^{\rm CP}(x,\mu),
\label{eq:fivexthirteen}
\end{equation}
one has
\begin{equation}
   \eta_\mu^{\rm CP}(x)=\cases{
   -U^{\rm CP}(x,i)\eta_i(\bar x-a\hat i)^*U^{\rm CP}(x,i)^{-1},&
   for $\mu=i$,\cr
   \eta_4(\bar x)^*,&for $\mu=4$.\cr}
\label{eq:fivexfourteen}
\end{equation}
Note that $(\eta_\mu^{\rm CP})^{\rm CP}(x)=\eta_\mu(x)$ corresponding to
$(U^{\rm CP})^{\rm CP}(x,\mu)=U(x,\mu)$. This definition of the
variation implies, in particular,
$\delta_\eta D(U^{\rm CP})=W\delta_\eta D(U)^TW^{-1}$.} If the measure
term is not invariant under CP, it then produces another unphysical
source of CP breaking as eq.~(\ref{eq:twoxtwentysix}) shows. In other
words, the requirement~(\ref{eq:fivextwelve}) eliminates an unnecessary
CP violation which may result from a wrong choice of the fermion measure
(which might be called ``fake CP anomaly''). Fortunately, it is always
possible to construct the CP invariant ideal measure term by the average
over CP~\cite{Luscher:1999du,Luscher:2000zd}:
\begin{eqnarray}
   &&{\mathfrak L}_\eta^{\{v,\overline v\}}[U;t]
   =a^4\sum_x\eta_\mu^a(x)j_\mu^a(x)[U;t]
\nonumber\\
   &&\to{1\over2}\biggl[a^4\sum_x\eta_\mu^a(x)j_\mu^a(x)[U;t]
   +a^4\sum_x\eta_\mu^{{\rm CP}a}(x)j_\mu^a(x)[U^{\rm CP};t]\biggr].
\label{eq:fivexfifteen}
\end{eqnarray}
This average is possible even if $U$ and $U^{\rm CP}$ belong to
different topological sectors $M$ and $M^{\rm CP}$, because the CP
operation defines a differentiable one-to-one onto-mapping from $M$
to~$M^{\rm CP}$. Then CP invariance of the measure term is ensured and
this is equivalent to eq.~(\ref{eq:fivexten}), and vice versa.

\section{Yukawa couplings}

It is straightforward to add the Yukawa coupling to the present
formulation.\footnote{It has been pointed out that a conflict with the
Majorana reduction in the presence of the Yukawa coupling and CP
symmetry are closely related to each
other~\cite{Fujikawa:2002is,Fujikawa:2001ka}.} By introducing the
right-handed Weyl fermion and the Higgs field, we set
\begin{eqnarray}
   S_{\rm F}&=&a^4\sum_x[
   \overline\psi_L(x)D\psi_L(x)+\overline\psi_R(x)D'\psi_R(x)
   +\overline\psi_L(x)\phi(x)\psi_R(x)
   +\overline\psi_R(x)\phi^\dagger(x)\psi_L(x)
\nonumber\\
   &&\qquad\qquad
   -\overline\psi_L(x)\eta_L(x)-\overline\eta_L(x)\psi_L(x)
   -\overline\psi_R(x)\eta_R(x)-\overline\eta_R(x)\psi_R(x)],
\label{eq:sixxone}
\end{eqnarray}
where
\begin{eqnarray}
   &&P_-^{(t)}\psi_L=\psi_L,\qquad
   \overline\psi_L\overline P_+^{(t)}=\overline\psi_L,
\nonumber\\
   &&P_+^{\prime(t)}\psi_R=\psi_R,\qquad
   \overline\psi_R\overline P_-^{\prime(t)}=\overline\psi_R.
\label{eq:sixxtwo}
\end{eqnarray}
We assume that the left-handed fermion~$\psi_L(x)$ belongs to the
representation~$R_L$ of the gauge group and the right-handed
fermion~$\psi_R(x)$ belongs to~$R_R$ (the Higgs field~$\phi(x)$
transforms as~$R_L\otimes(R_R)^*$). The gauge couplings in the Dirac
operator~$D$ ($D'$), and correspondingly in $P_-^{(t)}$
and~$\overline P_+^{(t)}$ ($P_+^{\prime(t)}$
and~$\overline P_-^{\prime(t)}$), are thus defined with respect to the
representation~$R_L$ ($R_R$).

We first note the relation\footnote{This formula assumes a perturbative
treatment of the Higgs coupling.}
\begin{eqnarray}
   &&Z_{\rm F}[U,\phi,\eta_L,\overline\eta_L,
   \eta_R,\overline\eta_R;t]
   =\int{\rm D}[\psi]{\rm D}[\overline\psi]\,e^{-S_{\rm F}}
\nonumber\\
   &&=\exp\biggl\{a^{-4}\sum_x\biggl[
   {\partial\over\partial\eta_L(x)}\overline P_+^{(t)}\phi(x)
   P_+^{\prime(t)}{\partial\over\partial\overline\eta_R(x)}
   +{\partial\over\partial\eta_R(x)}\overline P_-^{\prime(t)}
   \phi^\dagger(x)P_-^{(t)}
   {\partial\over\partial\overline\eta_L(x)}\biggr]\biggr\}
\nonumber\\
   &&\qquad\qquad\qquad\qquad\qquad\qquad\qquad\qquad\qquad\qquad
   \times Z_{\rm F}[U,0,\eta_L,\overline\eta_L,
   \eta_R,\overline\eta_R;t],
\label{eq:sixxthree}
\end{eqnarray}
because the fermion integration measure refers to neither source fields
nor the Higgs field. The generating functional without the Yukawa
coupling can be analyzed as before, and we have\footnote{It is
interesting to note that topologically non-trivial (i.e.,
$n_+-n_-\neq0$, $n_+'-n_-'\neq0$) sectors also contribute to fermion
number {\it non}-violating processes in the presence of the Yukawa
coupling. H.S. would like to thank Yoshio Kikukawa for a discussion on
this issue.}
\begin{eqnarray}
   &&Z_{\rm F}[U,0,\eta_L,\overline\eta_L,\eta_R,\overline\eta_R;t]
\nonumber\\
   &&=e^{i\theta[U;t]}\left({\Lambda\over a}\right)^N
   \prod^{n_-}\left[a^4\sum_x\overline\eta_L(x)\varphi_0^-(x)\right]
   \prod^{n_+}\left[a^4\sum_x\varphi_0^{+\dagger}(x)\eta_L(x)\right]
   \prod_{\lambda_j>0\atop\lambda_j\neq\Lambda}
   \left({\lambda_j\over a}\right)
\nonumber\\
   &&\qquad\qquad\qquad\qquad
   \times
   \exp\left[a^8\sum_{x,y}\overline\eta_L(x)G^{(t)}(x,y)\eta_L(y)
   \right]
\nonumber\\
   &&\qquad\qquad
   \times\left({\Lambda\over a}\right)^{\overline N'}
   \prod^{n_+'}
   \left[a^4\sum_x\overline\eta_R(x)\varphi_0^{\prime+}(x)\right]
   \prod^{n_-'}\left[a^4\sum_x\varphi_0^{\prime-\dagger}(x)
   \eta_R(x)\right]
   \prod_{\lambda_j'>0\atop\lambda_j'\neq\Lambda}
   \left({\lambda_j'\over a}\right)
\nonumber\\
   &&\qquad\qquad\qquad\qquad
   \times
   \exp\left[a^8\sum_{x,y}\overline\eta_R(x)G^{\prime(t)}(x,y)
   \eta_R(y)\right],
\label{eq:sixxfour}
\end{eqnarray}
as a product of left-handed and right-handed contributions. In this
expression, all quantities with the prime ($'$) are defined with respect
to $H'=a\gamma_5D'$ and
\begin{equation}
   D'G^{\prime(t)}(x,y)=\overline P_-^{\prime(t)}(x,y)
   -\sum^{n_-'}\varphi_0^{\prime-}(x)\varphi_0^{\prime-\dagger}(y).
\label{eq:sixxfive}
\end{equation}

By repeating the same arguments as before\footnote{Since the charge
conjugation flips the chirality as
\begin{equation}
   \overline\psi_RD'\psi_R
   =(\psi_R^TC)D_{R_R\to (R_R)^*}'(-C^{-1}\overline\psi_R^T),
\label{eq:sixxsix}
\end{equation}
and
\begin{equation}
   (\psi_R^TC)\overline P_{+,R_R\to(R_R)^*}^{(2-t)}=(\psi_R^TC),\qquad
   P_{-,R_R\to(R_R)^*}^{(2-t)}(-C^{-1}\overline\psi_R^T)
   =(-C^{-1}\overline\psi_R^T),
\label{eq:sixxseven}
\end{equation}
the right-handed fermion may be treated as the left-handed one,
belonging to the conjugate representation~$(R_R)^*$ (with the change
$t\to2-t$). In particular, the reconstruction theorem is applied with
trivial modifications.}
for~$Z_{\rm F}[U,0,\eta_L,\overline\eta_L,\eta_R,\overline\eta_R;t]$
we finally have, corresponding to eq.~(\ref{eq:fourxtwentyone}),
\begin{eqnarray}
   &&Z_{\rm F}[U^{\rm CP},\phi^\dagger,-W^{-1}\overline\eta_L^T,
   \eta_L^TW,-W^{-1}\overline\eta_R^T,\eta_R^TW;t]
\nonumber\\
   &&=e^{i\theta_M}\left({\Lambda\over a}\right)%
   ^{\overline N-N-\overline N'+N'}
\nonumber\\
   &&\times\exp\biggl\{a^{-4}\sum_x\biggl[
   {\partial\over\partial\eta_L(x)}\overline P_+^{(2-t)}\phi(x)
   P_+^{\prime(2-t)}{\partial\over\partial\overline\eta_R(x)}
   +{\partial\over\partial\eta_R(x)}\overline P_-^{\prime(2-t)}
   \phi^\dagger(x)P_-^{(2-t)}
   {\partial\over\partial\overline\eta_L(x)}\biggr]\biggr\}
\nonumber\\
   &&\qquad\qquad
   \times{\exp\biggl\{
   a^8\sum_{x,y}\biggl[\overline\eta_L(x)G^{(2-t)}(x,y)\eta_L(y)
   +\overline\eta_R(x)G^{\prime(2-t)}(x,y)\eta_R(y)\biggr]\biggr\}
   \over
   \exp\biggl\{
   a^8\sum_{x,y}\biggl[\overline\eta_L(x)G^{(t)}(x,y)\eta_L(y)
   +\overline\eta_R(x)G^{\prime(t)}(x,y)\eta_R(y)\biggr]\biggr\}}
\nonumber\\
   &&\qquad\qquad\qquad\qquad\qquad\qquad\qquad
   \times Z_{\rm F}[U,0,\eta_L,\overline\eta_L,
   \eta_R,\overline\eta_R;t].
\label{eq:sixxeight}
\end{eqnarray}
Thus we see that the effect of the CP breaking appears precisely in the
same places as before, except for the terms consisting of Yukawa
couplings connected by the propagators.\footnote{Note that the same
projection operator, $P_+^{\prime(2-t)}$ for example, appears in the
Yukawa vertex in the combination~$\phi(x)P_+^{\prime(2-t)}$ and in the
propagator of the Weyl fermion in the
combination~$P_+^{\prime(2-t)}/D'$. Consequently, it does not matter if
one says that CP is broken either by the propagator or by the Yukawa
vertex. When $R_L=R_R$, however, it is natural to combine $\psi_L$
and~$\psi_R$ into a Dirac fermion~$\psi$. In this case, the propagator
of~$\psi$ is manifestly CP invariant and the (chirally symmetric) Yukawa
vertex breaks CP.} However, when the Higgs field acquires the
expectation value, a completely new situation arises. Setting
$\phi(x)=v$, the fermion propagators read
\begin{eqnarray}
   &&{\langle\psi_L(x)\overline\psi_L(y)\rangle_{\rm F}\over
   \langle1\rangle_{\rm F}}
   =P_-^{(t)}{1\over D-v_+D^{\prime-1}v_-}\overline P_+^{(t)}(x,y),
\nonumber\\
   &&{\langle\psi_L(x)\overline\psi_R(y)\rangle_{\rm F}\over
   \langle1\rangle_{\rm F}}
   =P_-^{(t)}{1\over v_--D'v_+^{-1}D}\overline P_-^{\prime(t)}(x,y),
\nonumber\\
   &&{\langle\psi_R(x)\overline\psi_R(y)\rangle_{\rm F}\over
   \langle1\rangle_{\rm F}}
   =P_+^{\prime(t)}{1\over D'-v_-D^{-1}v_+}
   \overline P_-^{\prime(t)}(x,y),
\nonumber\\
   &&{\langle\psi_R(x)\overline\psi_L(y)\rangle_{\rm F}\over
   \langle1\rangle_{\rm F}}
   =P_+^{\prime(t)}{1\over v_+-Dv_-^{-1}D'}\overline P_+^{(t)}(x,y),
\label{eq:sixxnine}
\end{eqnarray}
where we have defined $v_+=\overline P_+^{(t)}vP_+^{\prime(t)}$
and~$v_-=\overline P_-^{\prime(t)}v^\dagger P_-^{(t)}$. One thus sees
that this time the change $t\to 2-t$ produces {\it non-local\/}
differences in the propagator. For example, in
$\langle\psi_L(x)\overline\psi_R(y)\rangle_{\rm F}/%
\langle1\rangle_{\rm F}$, the difference
$\overline P_-^{\prime(2-t)}-\overline P_-^{\prime(t)}\propto%
a(t-1)D'\gamma_5f(H^{\prime2})$ cannot cancel the
denominator~$v_--D'v_+^{-1}D$, leaving a non-local difference.%
\footnote{Although the kernel $1/(v_--D'v_+^{-1}D)(x,y)$ decays
exponentially as $|x-y|\to\infty$, this cannot be regarded as local; the
decaying rate in the lattice unit is
$\sim 1/(\sqrt{vv^\dagger}a)\to\infty$ in the continuum limit~$a\to0$,
because $vv^\dagger$ is kept fixed in this limit (i.e., $vv^\dagger$ has
the physical mass scale).} Though we expect that this breaking, even if
it is non-local, will eventually be removed in a suitable continuum
limit, a more careful study is required to confirm this
expectation.\footnote{This non-local CP breaking will persist for a
non-perturbative treatment of the Higgs coupling, though a detailed
analysis remains to be performed.} (If one forms the free Dirac-type
propagator, the CP breaking does not appear in the propagator. This
means that the coupling of chiral gauge fields induces CP breaking.)

\section{Conclusion and discussion}

In this paper we have analyzed the possible implications of CP breaking
in lattice chiral gauge theory, which is a result of the very definition
of chirality~(\ref{eq:onexone}) for the Ginsparg-Wilson
operator~\cite{Hasenfratz:2001bz}. This CP breaking is known to be
directly related to the basic notions of locality and the absence of
species doubling in the Ginsparg-Wilson operator~\cite{Fujikawa:2002is}.
Although the non-perturbative construction of the ideal path integral
measure for non-abelian chiral gauge theories has not been established
yet, we analyzed the CP transformation properties of the path integral
measure on the basis of a working ansatz. Our conclusion is that there
exists no ``CP anomaly'' arising from the path integral measure. The
breaking of CP is thus limited to the explicit breaking in the action of
chiral gauge theory, and it basically appears in the fermion propagator.
When the Higgs field has no vacuum expectation value or in pure chiral
gauge theory without the Higgs field, it emerges as as an (almost)
contact term. In the presence of the Higgs expectation value, however,
the breaking becomes intrinsically non-local. We expect that these
breakings in the propagator, either local or non-local, do not survive
in a suitable continuum limit, but a more careful analysis is required
to make a definite conclusion of this issue.

In this paper, we identified where the effect of CP breaking in the
present lattice formulation appears for generic chiral gauge theories.
When more general composite operators are considered, further
complications associated to this effect could arise. As an example, we
comment on the computation of the kaon $B$~parameter, $B_K$~(see, for
example, refs.~\cite{Lellouch:2000bm,Martinelli:2001yn}). The following
matrix element of the effective weak Hamiltonian is then relevant:
\begin{equation}
   \langle\overline K^0|\,\overline s\gamma_5\Gamma_5
   \gamma_\mu{1-\gamma_5\over2}\gamma_5\Gamma_5 d\,\,
   \overline s\gamma_5\Gamma_5
   \gamma_\mu{1-\gamma_5\over2}\gamma_5\Gamma_5 d\,|K^0\rangle,
\label{eq:sevenxone}
\end{equation}
where we have adopted the $O(a)$~improved
operator~\cite{Capitani:1999uz}. Since the gauge action and the
Ginsparg-Wilson action in QCD are invariant under CP,
eq.~(\ref{eq:sixxone}) is equal to its CP transformation
\begin{equation}
   \langle K^0|\,\overline d\gamma_5\Gamma_5
   \gamma_\mu{1-\gamma_5\over2}\gamma_5\Gamma_5 s\,\,
   \overline d\gamma_5\Gamma_5
   \gamma_\mu{1-\gamma_5\over2}\gamma_5\Gamma_5 s\,
   |\overline K^0\rangle.
\label{eq:sevenxtwo}
\end{equation}
which coincides with the lattice transcription of the {\it naive\/} CP
transformation of~eq.~(\ref{eq:sevenxone}). This shows that the
$O(a)$~improvement in ref.~\cite{Capitani:1999uz}, which eliminates
$O(a)$ chiral symmetry breakings (in the sense of continuum theory),
maintains the desired behavior of the amplitude (\ref{eq:sevenxone})
under CP.

{}From a view point of the present analysis of CP symmetry, however, the
above $O(a)$~improved expression of the effective Hamiltonian, if
applied to off-shell amplitudes, is not completely
satisfactory.\footnote{For on-shell amplitudes such as in
eq.~(\ref{eq:sevenxone}), the amplitude is reduced to the one in
continuum $V-A$ theory if the equations of motion for quarks are used.
In this sense, eq.~(\ref{eq:sevenxone}) and other schemes are
consistent. We thank Martin L\"uscher for bringing this fact to our
attention.} For example, one can confirm that the right-handed component
of the $\overline s(1-\gamma_5)/2$ quark\footnote{The right-handed
component of an anti-quark does not couple to the $W$~boson either in
continuum theory or in the standard lattice formulation with the overlap
operator.} contributes to the above {\it weak\/} effective Hamiltonian
in eq.~(\ref{eq:sevenxone}), if applied to off-shell Green's functions.
Although this is the order $O(a)$ effect, this breaks $\SU(2)_L$ gauge
symmetry of electroweak interactions. This illustrates that great care
need to be exercised in the analysis of lattice chiral symmetry and CP
invariance.

\appendix

\section{C, P and CP}
We adopt the following convention of $\gamma$-matrices:
\begin{eqnarray}
   &&\{\gamma_\mu,\gamma_\nu\}=2\delta_{\mu\nu},\qquad
   \gamma_\mu^\dagger=\gamma_\mu,\qquad
   \gamma_5=-\gamma_1\gamma_2\gamma_3\gamma_4=\gamma_5^\dagger,
\nonumber\\
   &&\gamma_1^T=-\gamma_1,\qquad\gamma_2^T=\gamma_2,\qquad
   \gamma_3^T=-\gamma_3,\qquad\gamma_4^T=\gamma_4,\qquad
   \gamma_5^T=\gamma_5.
\end{eqnarray}

The charge conjugation is defined by
\begin{eqnarray}
   &&\psi(x)\to-C^{-1}\overline\psi^T(x),\qquad
   \overline\psi(x)\to\psi^T(x)C,
\nonumber\\
   &&U(x,\mu)\to U^{\rm C}(x,\mu)=U(x,\mu)^*,
\end{eqnarray}
where the charge conjugation matrix~$C=\gamma_2\gamma_4$ satisfies
\begin{equation}
   C^\dagger C=1,\qquad C^T=-C,\qquad C\gamma_\mu C^{-1}=-\gamma_\mu^T,
   \qquad C\gamma_5C^{-1}=\gamma_5^T.
\end{equation}
Under this transformation, the kernel of Dirac operator transforms
as\footnote{We assume that the basic building block of the Dirac
operator is the Wilson-Dirac operator.}
\begin{equation}
   D(U^{\rm C})(x,y)=C^{-1}D(U)(x,y)^TC,
\end{equation}
where the transpose operation~$T$ acts not only on the matrices involved
but also on the arguments as $(x,y)\to(y,x)$.

The parity transformation is defined by
\begin{eqnarray}
   &&\psi(x)\to\gamma_4\psi(\bar x),\qquad
   \overline\psi(x)\to\overline\psi(\bar x)\gamma_4,
\nonumber\\
   &&U(x,\mu)\to U^{\rm P}(x,\mu)=\cases{
   U(\bar x-a\hat i,i)^{-1},&for $\mu=i$,\cr
   U(\bar x,4),& for $\mu=4$,\cr}
\end{eqnarray}
where
\begin{equation}
   \bar x=(-x_i,x_4),
\end{equation}
for $x=(x_i,x_4)$ ($i=1$, $2$, $3$). Under this,
\begin{equation}
   D(U^{\rm P})(x,y)=\gamma_4D(U)(\bar x,\bar y)\gamma_4.
\end{equation}

Finally, we define the CP transformation as
\begin{eqnarray}
   &&\psi(x)\to-W^{-1}\overline\psi^T(\bar x),\qquad
   \overline\psi(x)\to\psi^T(\bar x)W
\nonumber\\
   &&U(x,\mu)\to U^{\rm CP}(x,\mu)=\cases{
   {U(\bar x-a\hat i,i)^{-1}}^*,&for $\mu=i$,\cr
   U(\bar x,4)^*,& for $\mu=4$,\cr}
\end{eqnarray}
where
\begin{eqnarray}
   &&W=\gamma_2,\qquad W^\dagger W=1,
\nonumber\\
   &&W\gamma_\mu W^{-1}
   =\cases{\gamma_i^T,&for $\mu=i$,\cr
   -\gamma_4^T,& for $\mu=4$,\cr}\qquad
   W\gamma_5 W^{-1}=-\gamma_5^T,
\end{eqnarray}
and thus CP acts on the plaquette variables $U(x,\mu,\nu)$ as
($\varphi(\bar x-a\hat i-a\hat j,i,j)%
=U(\bar x-a\hat i-a\hat j,j)^*U(\bar x-a\hat i,i)^*$)
\begin{eqnarray}
   &&U(x,i,j)
\nonumber\\
   &&\to U^{\rm CP}(x,i,j)
   =\varphi(\bar x-a\hat i-a\hat j,i,j)^{-1}
   U(\bar x-a\hat i-a\hat j,i,j)^*
   \varphi(\bar x-a\hat i-a\hat j,i,j),
\nonumber\\
   &&U(x,i,4)\to U^{\rm CP}(x,i,4)
   ={U(\bar x-a\hat i,i)^*}^{-1}
   {U(\bar x-a\hat i,i,4)^{-1}}^*U(\bar x-a\hat i,i)^*.
\end{eqnarray}
Under this transformation, we have
\begin{equation}
   D(U^{\rm CP})(x,y)=W^{-1}D(U)(\bar x,\bar y)^TW.
\end{equation}

\section{Eigenvalue problem of $H^2$}

To consider the eigenvalue problem of $H^2$~(\ref{eq:threexone}), it is
better to consider first
\begin{equation}
   H\varphi_n(x)=\lambda_n\varphi_n(x),\qquad
   (\varphi_n,\varphi_m)=\delta_{nm}.
\label{eq:axone}
\end{equation}
We note
\begin{equation}
   H\Gamma_5\varphi_n(x)=-\Gamma_5H\varphi_n(x)
   =-\lambda_n\Gamma_5\varphi_n(x),
\end{equation}
and
\begin{equation}
   (\Gamma_5\varphi_n,\Gamma_5\varphi_m)
   =[1-\lambda_n^2f^2(\lambda_n^2)]\delta_{nm}.
\end{equation}
These relations show that eigenfunctions with $\lambda_n\neq0$ (when
$\lambda_n=0$, $\varphi_0(x)$ and $\Gamma_5\varphi_0(x)$ are not
necessarily linear-independent) and $\lambda_nf(\lambda_n^2)\neq\pm1$
come in pairs as $\lambda_n$ and~$-\lambda_n$.

We can thus classify eigenfunctions in eq.~(\ref{eq:axone}) as follows:

(i) $\lambda_n=0$ ($H\varphi_0(x)=0$). For this
\begin{equation}
   H{1\pm\gamma_5\over2}\varphi_0(x)=H{1\pm\Gamma_5\over2}\varphi_0(x)
   ={1\mp\Gamma_5\over2}H\varphi_0(x)=0,
\end{equation}
so we may impose the chirality on $\varphi_0(x)$ as
\begin{equation}
   \gamma_5\varphi_0^\pm(x)=\Gamma_5\varphi_0^\pm(x)
   =\pm\varphi_0^\pm(x).
\end{equation}
We denote the number of $\varphi_0^+(x)$ ($\varphi_0^-(x)$) as $n_+$
($n_-$).

(ii) $\lambda_n\neq0$ and $\lambda_nf(\lambda_n^2)\neq\pm1$. As shown
above,
\begin{equation}
   H\varphi_n(x)=\lambda_n\varphi_n(x),\qquad
   H\widetilde\varphi_n(x)=-\lambda_n\widetilde\varphi_n(x),
\end{equation}
where
\begin{equation}
   \widetilde\varphi_n(x)={1\over\sqrt{1-\lambda_n^2f^2(\lambda_n^2)}}
   \Gamma_5\varphi_n(x).
\end{equation}
We have
\begin{equation}
   \Gamma_5\varphi_n(x)=\sqrt{1-\lambda_n^2f^2(\lambda_n^2)}
   \widetilde\varphi_n(x),\qquad
   \Gamma_5\widetilde\varphi_n(x)=\sqrt{1-\lambda_n^2f^2(\lambda_n^2)}
   \varphi_n(x),
\end{equation}
and
\begin{equation}
   \gamma_5\pmatrix{\varphi_n\cr\widetilde\varphi_n\cr}
   =\pmatrix{\lambda_nf(\lambda_n^2)
             &\sqrt{1-\lambda_n^2f^2(\lambda_n^2)}\cr
             \sqrt{1-\lambda_n^2f^2(\lambda_n^2)}
             &-\lambda_nf(\lambda_n^2)\cr}
   \pmatrix{\varphi_n\cr\widetilde\varphi_n\cr}.
\end{equation}

(iii) $\lambda_nf(\lambda_n^2)=\pm1$, or
\begin{equation}
   H\Psi_\pm(x)=\pm\Lambda\Psi_\pm(x),\qquad\Lambda f(\Lambda^2)=1.
\end{equation}
We see
\begin{equation}
   \Gamma_5\Psi_\pm(x)=0,
\end{equation}
and
\begin{equation}
   \gamma_5\Psi_\pm(x)=\pm\Lambda f(\Lambda^2)\Psi_\pm(x)
   =\pm\Psi_\pm(x).
\end{equation}
We denote the number of $\Psi_+(x)$ ($\Psi_-(x)$) as $N_+$ ($N_-$).

As the application of the above relations, we can establish the index
theorem~\cite{Hasenfratz:1998ri}. Noting $(\varphi_n,\Gamma_5\varphi_n)
=(\widetilde\varphi_n,\Gamma_5\widetilde\varphi_n)=0$ for modes with
$\lambda_n\neq0$, we see
\begin{equation}
   \Tr\Gamma_5=n_+-n_-.
\end{equation}
Since $\Gamma_5$ depends smoothly on the gauge field within the space of
admissible configurations~(\ref{eq:twoxthirteen}), the integer $n_+-n_-$
is a constant in a connected component of the space of admissible
configurations; $n_+-n_-$ thus provides a topological characterization
of the gauge field configuration, i.e., the index.

Next, using above relations, we have
\begin{equation}
   \Tr\gamma_5=n_+-n_-+N_+-N_-.
\end{equation}
If we further note $\Tr\gamma_5=0$ on the lattice, we have the chirality
sum rule~\cite{Chiu:1998bh,Fujikawa:1999ku}
\begin{equation}
   n_+-n_-+N_+-N_-=0.
\end{equation}

Going back to our original problem (\ref{eq:threexone}), it is obvious
that eigenfunctions are given by the above eigenfunctions of $H$, by
identifying $\varphi_j\leftrightarrow\varphi_n$ and
$\lambda_j\leftrightarrow|\lambda_n|$ (so $\lambda_j$ is doubly
degenerated). To find appropriate components for
eq.~(\ref{eq:threextwo}), we have to know the action of chiral
projectors on these eigenfunctions $\varphi_n$. From above, we have

(i)~For zero modes, we simply have
\begin{equation}
   \gamma_5^{(t)}\varphi_0^\pm=\gamma_5\varphi_0^\pm=\pm\varphi_0^\pm,
\end{equation}
and thus
\begin{equation}
   P_\pm^{(t)}\varphi_0^\pm=\pm\varphi_0^\pm.
\end{equation}

(ii) For modes with $\lambda_n\neq0$ and $\lambda_n\neq\pm\Lambda$,
\begin{eqnarray}
   &&\gamma_5^{(t)}\pmatrix{\varphi_n\cr\widetilde\varphi_n\cr}
\\
   &&={1\over\sqrt{1+t(t-2)\lambda_n^2f^2(\lambda_n^2)}}
   \pmatrix{(1-t)\lambda_nf(\lambda_n^2)
             &\sqrt{1-\lambda_n^2f^2(\lambda_n^2)}\cr
             \sqrt{1-\lambda_n^2f^2(\lambda_n^2)}
             &-(1-t)\lambda_nf(\lambda_n^2)\cr}
   \pmatrix{\varphi_n\cr\widetilde\varphi_n\cr}.
\nonumber
\end{eqnarray}
Since this is a traceless matrix whose determinant is~$-1$, the
eigenvalues of $\gamma_5^{(t)}$ in this subspace are $+1$ and $-1$. This
shows that one linear combination of $\varphi_n$ and
$\widetilde\varphi_n$ is annihilated by $P_+^{(t)}$ and the orthogonal
combination is annihilated by $P_-^{(t)}$. Therefore we can take
suitable linear combinations of $\varphi_n$ and~$\widetilde\varphi_n$
such that
\begin{equation}
   P_\pm^{(t)}\varphi_n^\pm(x)=\varphi_n^\pm(x),\qquad
   (\varphi_n^\pm,\varphi_m^\pm)=\delta_{nm}.
\end{equation}

(iii) For the modes with $\lambda_n=\pm\Lambda$, we have
\begin{equation}
   \gamma_5^{(t)}\Psi_\pm(x)=\cases{\mp\Psi_\pm(x),&for $t>1$,\cr
   \pm\Psi_\pm(x),&for $t<1$.\cr}
\end{equation}

\listoftables		
\listoffigures		

\end{document}